\documentclass[11pt,preprint]{aastex}

\begin{document}
\newcommand{\myvert}[2]{$\begin{array}{c} #1 \\  #2\end{array}$}
\shorttitle{TW Hya Disk Kinematics}

\shortauthors{Rosenfeld et al.}

\title{Kinematics of the CO Gas in the Inner Regions of the TW Hya Disk}

\author{Katherine A. Rosenfeld\altaffilmark{1}, Chunhua Qi\altaffilmark{1}, Sean M. Andrews\altaffilmark{1}, David J. Wilner\altaffilmark{1}, Stuartt A. Corder\altaffilmark{2}, \\ C. P. Dullemond\altaffilmark{3}, Shin-Yi Lin\altaffilmark{4}, A. M. Hughes\altaffilmark{5}, Paola D'Alessio\altaffilmark{6}, and P. T. P. Ho\altaffilmark{7}}

\altaffiltext{1}{Harvard-Smithsonian Center for Astrophysics, 60 Garden Street, Cambridge, MA 02138}
\altaffiltext{2}{National Radio Astronomy Observatory, 520 Edgemont Road, Charlottesville, VA 22903}
\altaffiltext{3}{Institut f{\"{u}}r Theoretische Astrophysik, Universit{\"{a}}t Heidelberg, Albert-Ueberle-Str.~2, 69120 Heidelberg, Germany}
\altaffiltext{4}{Deparment of Physics, University of California San Diego, 9500 Gilman Drive, La Jolla, CA 92093}
\altaffiltext{5}{University of California at Berkeley, Department of Astronomy, 601 Campbell Hall, Berkeley, CA 94720}
\altaffiltext{6}{Centro de Radioastronom{\'{\i}}a y Astrof{\'{\i}}sica, Universidad Nacional Aut{\'{o}}noma de M{\'{e}}xico, Apartado Postal 72-3 (Xangari), 58089 Morelia, Michoac{\'{a}}n, Mexico}
\altaffiltext{7}{Academia Sinica Institute of Astronomy and Astrophysics, P.O. Box 23-141, Taipei 106, Taiwan}

\begin{abstract}
We present a detailed analysis of the spatially and spectrally resolved 
$^{12}$CO $J$=2$-$1 and $J$=3$-$2 emission lines from the TW Hya circumstellar
disk, based on science verification data from the Atacama Large 
Millimeter/Submillimeter Array (ALMA).  These lines exhibit substantial 
emission in their high-velocity wings (with projected velocities out to 
2.1\,km s$^{-1}$, corresponding to intrinsic orbital velocities $>$20\,km 
s$^{-1}$) that trace molecular gas as close as 2\,AU from the central star.  
However, we are not able to reproduce the intensity of these wings and the 
general spatio-kinematic pattern of the lines with simple models for the disk 
structure and kinematics.  Using three-dimensional non-local thermodynamic 
equilibrium molecular excitation and radiative transfer calculations, we 
construct some alternative models that successfully account for these features 
by modifying either (1) the temperature structure of the inner disk (inside 
the dust-depleted disk cavity; $r < 4$\,AU); (2) the intrinsic (Keplerian) disk 
velocity field; or (3) the distribution of disk inclination angles (a warp).  
The latter approach is particularly compelling because a representative warped 
disk model qualitatively reproduces the observed azimuthal modulation of 
optical light scattered off the disk surface.  In any model scenario, the ALMA 
data clearly require a substantial molecular gas reservoir located inside the 
region where dust optical depths are known to be substantially diminished in 
the TW Hya disk, in agreement with previous studies based on infrared 
spectroscopy.  The results from these updated model prescriptions are discussed 
in terms of their potential physical origins, which might include dynamical 
perturbations from a low-mass companion with an orbital separation of a few AU.
\end{abstract}

\section{Introduction}

In the standard model for star formation, the collapse of a slowly rotating 
molecular cloud core at least partially conserves its angular momentum by 
forming a compact, flattened disk that channels mass onto a central protostar 
\citep{cassen81,terebey84}.  Once the remnant core material is accreted or 
dispersed, the gravitational potential of the star determines the kinematic 
properties of its disk.  At that time, the intrinsic disk velocity field should 
be described well by a Keplerian pattern of differential rotation in circular 
orbits, $v_{\phi}(r) = v_k = \sqrt{G M/r}$, where $\phi$ is the azimuthal 
direction in a polar coordinate system, $G$ is the gravitational constant, $r$ 
is the distance from the star, and $M$ is the mass enclosed within $r$.  In 
most cases, the disk-to-star mass ratio is low enough that $M \approx 
M_{\ast}$ is a reasonable approximation.  Imaging spectroscopy data taken with 
millimeter interferometers first verified these signatures of disk rotation 
using the optically thick emission lines of the carbon monoxide (CO) molecule 
\citep{koerner93,dutrey94,koerner95,mannings97}.  Subsequent work demonstrated 
that the rotation patterns were indeed Keplerian ($v_{\phi} \propto r^{-0.5}$) 
and could be used to make dynamical estimates of central star masses 
\citep{guilloteau98,dutrey98,simon00}.

In practice, spectral imaging observations measure the intrinsic velocity field 
projected onto the sky, $v_{\rm obs} = v_{\phi} \sin{i}$, where $i$ is the 
angle between the disk rotation axis and the observed line of sight (e.g., $i = 
0\degr$ corresponds to face-on).  A Keplerian velocity field can be exploited 
to probe gas at small radii in the disk, even if the angular resolution of the 
observations is relatively limited \citep[e.g.,][]{dutrey08}.  This kind of 
physically motivated super-resolution can be achieved with a focus on the 
intensities (and ideally morphologies) in the high-velocity wings of emission 
lines.  For optically thick lines like those produced by the rotational 
transitions of CO, the line strength depends on the product of the gas 
temperature and emitting area \citep[e.g.,][]{beckwith93}.  Since the projected 
emitting areas are typically small in the inner disk, emission in the line 
wings is generally weak; often well below the sensitivity thresholds for most 
millimeter-wave interferometers.   

To further complicate issues, there are theoretical mechanisms and disk 
properties that could modify this simple prescription for interpreting the 
projected disk velocity field.  For example, in a very massive disk, 
gravitational instabilities could excite spiral waves that drive significant 
(radial) streaming motions.  This was suggested as a potential explanation for 
the complex kinematic environment of the AB Aur disk \citep{lin06}, although 
others have associated its (possibly) non-Keplerian motions with envelope 
contamination \citep{pietu05,corder05}.  Gas pressure gradients also lead to 
deviations from Keplerian orbits: the sub-Keplerian shift expected from thermal 
(hydrostatic) pressure is too small to observe directly 
\citep{weidenschilling77}, but magnetic pressure might have a more substantial 
influence over gas motions \citep[e.g.,][]{shu07,shu08}.  Alternatively, a disk 
might {\it appear} to have non-Keplerian motions (even if $v_{\phi} = v_k$ 
exactly) if its line-of-sight orientation ($i$) varies with radius -- that is, 
if the disk structure is warped.  

Given that emission line wings are typically expected to be weak, disentangling 
these subtle (hypothetical) modifications to a simple projected velocity field 
from the intrinsic properties of the inner disk makes for a difficult task in 
data analysis.  Nevertheless, the potential for this unique access to the 
spatio-kinematic properties and physical conditions of the gas in the innermost 
regions of protoplanetary disks is extraordinarily compelling.  Perhaps the 
best opportunity to explore these features is with the massive, gas-rich disk 
around the nearest \citep[$d \approx 54$\,pc;][]{vanleeuwen07} classical T 
Tauri star, TW Hya.  Aside from the enhanced sensitivity and spatial resolution 
afforded by its proximity, the TW Hya disk coincidentally has the added 
benefits of a nearly face-on viewing geometry \citep[$i \approx 
6$-7\degr;][]{krist00,qi04,hughes11}.  This means that the line wing emission 
generated by gas in the inner regions of the TW Hya disk has relatively small 
observed Doppler shifts from the systemic velocity.  Because the low disk 
inclination angle minimizes radial projection effects, subtle departures from a 
simple (unwarped, Keplerian) model for the projected velocity field should be 
more easily recognizable observationally.  

In this article, we take advantage of the dramatically improved sensitivity 
now available in the Science Verification (SV) data products from the Atacama 
Large Millimeter/Submillimeter Array (ALMA) to help characterize the 
spatio-kinematic morphologies of the spatially and spectrally resolved CO 
$J$=2$-$1 and $J$=3$-$2 emission lines from the TW Hya circumstellar disk.  A 
brief overview of the data and its calibration are presented in \S 2.  The data 
is investigated in detail in \S 3, using emission line radiative transfer 
calculations and considering various toy models for the disk properties.  The 
results are synthesized in the larger context of the structure and kinematics 
in the inner regions of the TW Hya disk in \S 4, and summarized in \S 5.

\section{Data and Calibration}

The TW Hya disk was observed with nine 12-m elements of ALMA as part of the 
commissioning and SV effort during construction.  These data were obtained in 
two (contiguous) $\sim$4.5-hour observations, using receiver bands 6 and 7 on 
2011 April 20 and 22, respectively, with antenna spacings providing baseline 
lengths of $\sim$20-100\,m (one antenna was not available for band 7 
observations at this time).  The ALMA correlator was configured to 
simultaneously observe four 0.5\,GHz-wide spectral windows (two in each 
sideband) with a 122\,kHz channel spacing in each window.  One of those windows 
was centered near the $^{12}$CO $J$=2$-$1 line (230.538\,GHz) in band 6 and the 
$^{12}$CO $J$=3$-$2 line (345.796\,GHz) in band 7, providing a velocity spacing 
of 0.16 and 0.11\,km s$^{-1}$, respectively.  The ALMA correlator Hanning 
smoothes the data, making the effective spectral resolution coarser ($\sim$0.3 
and 0.2\,km s$^{-1}$ in bands 6 and 7).  Two other spectral windows were 
designed to provide $\sim$1\,GHz of continuum bandwidth in each receiver band; 
the remaining windows contained weaker spectral lines that are not of interest 
here.  Observations cycled between TW Hya and the nearby complex gain 
calibrator J1037-295 every $\sim$10\,minutes.  The bright quasar 3C 279 was 
observed as a bandpass calibrator, and brief observations were made of Titan 
for absolute flux calibration.

These data were calibrated in the {\tt CASA} software package (v3.3), carefully 
following the detailed processing scripts kindly provided by the ALMA science 
verification team.  Since those instructions, along with fully calibrated 
measurement sets, are publicly available online\footnote{\url{https://almascience.nrao.edu/alma-data/science-verification}.}, we do not repeat the details 
here.  The continuum-subtracted, calibrated spectral visibilities for each CO 
emission line were Fourier inverted (assuming natural weighting) with modest 
spectral averaging, deconvolved with the {\tt CLEAN} algorithm using a simple 
polygon mask applied to all channels, and restored with a synthesized beam of 
dimensions $2\farcs8\times2\farcs4$ (at a P.A.~of 44\degr) for $J$=2$-$1 and 
$1\farcs7\times1\farcs5$ (P.A. = 22\degr) for $J$=3$-$2.  We utilized two 
different channel spacings: (1) ``native", meaning 0.20 and 0.12\,km s$^{-1}$ 
for $J$=2$-$1 and $J$=3$-$2, respectively, to highlight the key kinematic 
features (note that this over-samples the true spectral resolution; this is the 
averaging advocated in the ALMA calibration tutorials), and (2) ``binned", 
meaning 0.35\,km s$^{-1}$ for both lines, to simplify comparisons and sample 
models closer to the true Hanning-smoothed velocity resolution.  The resulting 
spectral image cubes are indistinguishable from the reference datasets provided 
by the science verification team, with one small distinction: we elected to 
force the reduced spectral datasets to have a channel precisely centered on the 
known systemic LSR velocity of TW Hya, $+$2.87\,km s$^{-1}$.  

\begin{figure}[t!]
\epsscale{0.95}
\plotone{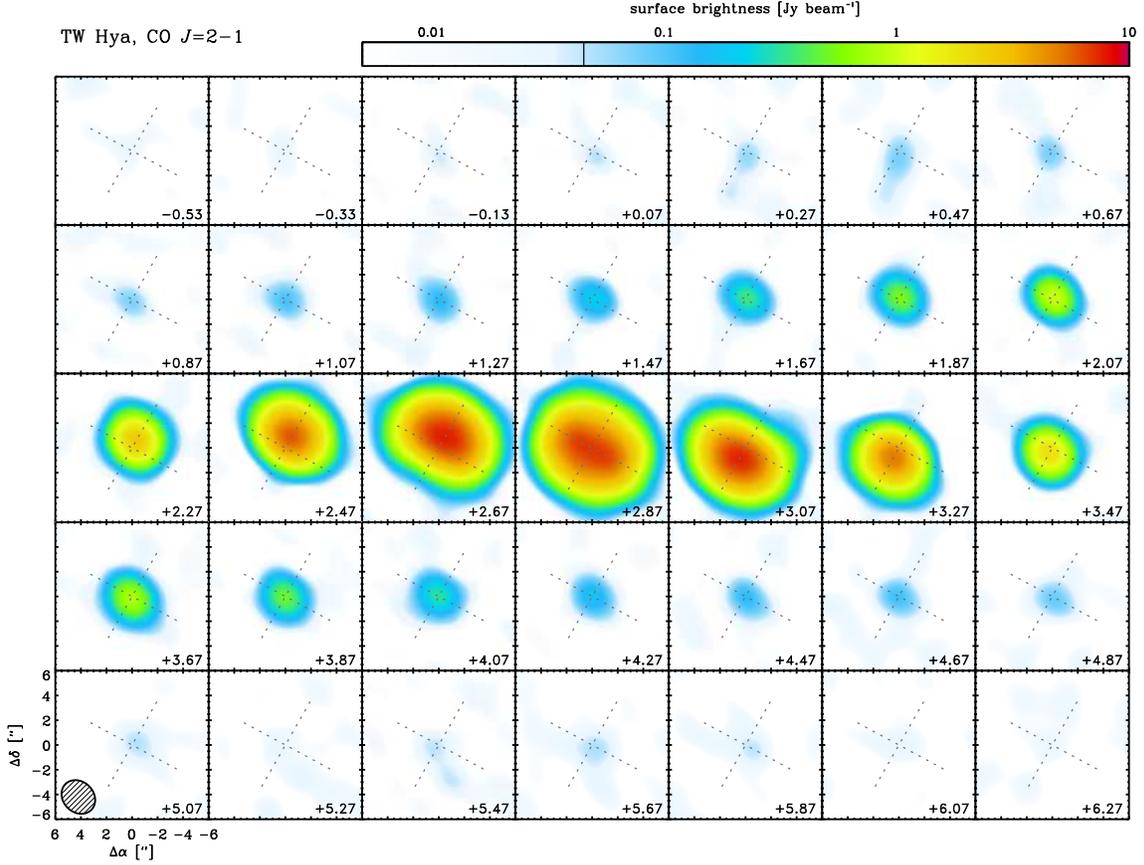}
\figcaption{Channel maps of the CO $J$=2$-$1 line emission from the TW Hya 
disk, with a ``native" velocity-spacing of 0.20\,km s$^{-1}$.  A logarithmic 
color-scale bar ({\it top}) indicates the line intensities, with a vertical 
line representing the 5\,$\sigma$ level (see Table \ref{tab:data}).  Each panel 
is 12\arcsec\ on a side, corresponding to a projected physical scale of 
$\sim$650\,AU.  Cross-hatches mark the stellar position and orientation of the 
disk major axis (151\degr\ east of north).  The synthesized beam dimensions are 
drawn in the bottom left panel; LSR velocities (in km s$^{-1}$) are marked in 
the bottom right of each panel.  \label{fig:co21}}
\end{figure}

\begin{figure}[ht!]
\epsscale{0.95}
\plotone{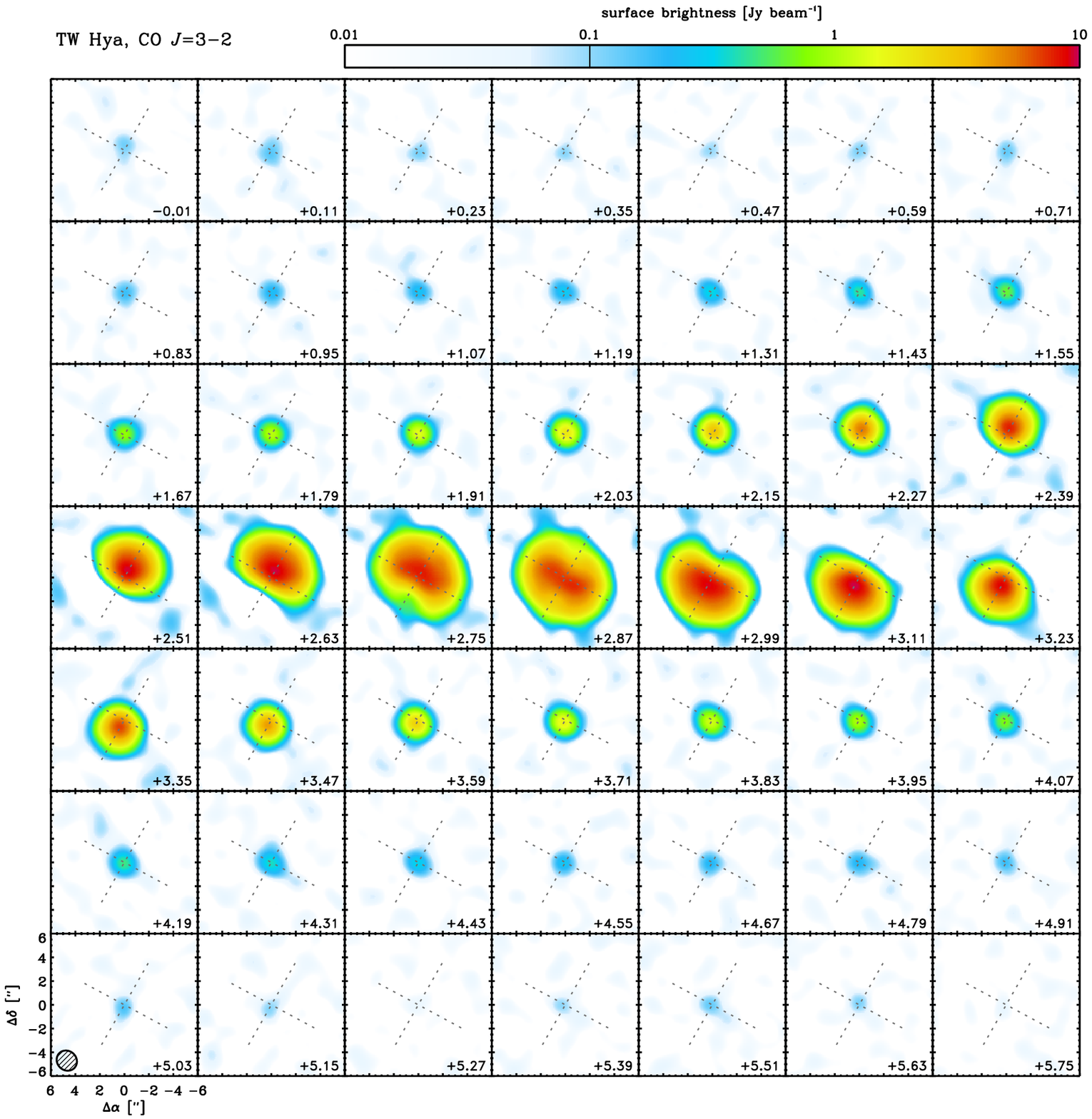}
\figcaption{Same as Figure \ref{fig:co21}, but for the CO $J$=3$-$2 line.
\label{fig:co32}}
\end{figure}

The CO $J$=2$-$1 and $J$=3$-$2 line emission from the TW Hya disk are shown as 
channel maps in Figures \ref{fig:co21} and \ref{fig:co32}, respectively, with a 
logarithmic color-scale stretch to emphasize emission over a wide dynamic 
range (a scale bar is marked on the top of each figure).  Measurements of the 
RMS noise levels, peak and integrated line intensities for each transition are 
compiled in Table \ref{tab:data}.  The channel maps clearly show a resolved 
pattern of rotation, from the blueshifted northwest to the redshifted southeast 
(with a major axis position angle of 151\degr), around the systemic velocity 
and centered at the stellar position \citep{qi04,qi06}.  While these are not 
the highest spectral \citep[see][]{hughes11} or spatial 
\citep[see][]{andrews12} resolution views of CO emission from the TW Hya disk, 
they are by far the most sensitive ($\sim$10$\times$ higher S/N).  This 
enhanced sensitivity reveals a surprising spectral feature: significant 
($>$5\,$\sigma$) emission in the wings of both CO transitions out to projected 
velocities at least $\pm$2.1\,km s$^{-1}$ from the line centers.

We will demonstrate in \S 3 that these line wings are primarily generated well 
inside a radius of $\sim$5\,AU in the TW Hya disk and, coupled with the 
emission morphology near the systemic velocity, they are difficult to reproduce 
for a disk in Keplerian rotation with a nearly face-on viewing geometry and 
smoothly-varying temperature profile.  Because of the value of this {\it 
relatively} faint emission at high velocities in what follows, we conducted a 
variety of tests to ensure that these line wings were not artificially 
generated in the calibration process.  First, we verified that the disk 
emission is not spectrally coincident with CO features in the Earth's 
atmosphere.  Weak and narrow CO absorption ($\sim$1\,MHz wide; $\sim$10-20\%\ 
depths) was noted and flagged in the observations of the bandpass calibrator 3C 
279: however, they are located $\gtrsim$20\,MHz from the TW Hya line center, 
and are therefore well outside the spectral range of interest here.  A second 
test focused on the remote possibility that the application of the 
frequency-dependent system temperatures ($T_{\rm sys}$) could have artificially 
broadened the lines.  In these data, a coarse measure of the spectral $T_{\rm 
sys}$ behavior was determined directly from observations of TW Hya.  The 
concern is that the bright CO emission from the disk might produce a weak, but 
broad feature in the $T_{\rm sys}$ profile that would then be propagated into 
the data as faint emission wings.  To test this, we re-calibrated the data 
instead with a spectrally flat, band-averaged $T_{\rm sys}$ correction.  This 
had no discernable effect on the results; the observed line wings persist at 
the same intensity levels.  Finally, we investigated whether the continuum 
subtraction process could be responsible.  A wide range of techniques were 
attempted, varying the polynomial order and channel ranges employed to 
determine the underlying continuum level on either side of the emission line.  
In all cases, we found no significant difference in the resulting line wings 
(indeed, they are even present without any continuum subtraction applied).  To 
summarize, we find no obvious artificial mechanism in the data acquisition or 
calibration procedures that can produce these substantial emission line wings 
with the same velocity-widths in two CO transitions observed at different times 
and with different receivers and spectral setups.

\section{Analysis}

Having presented the ALMA data (see Figures \ref{fig:co21} and \ref{fig:co32}), 
we now turn to its interpretation with a special emphasis on the resolved 
spatio-kinematic patterns of the CO emission lines.  First, we briefly describe 
a generic framework for producing models of the CO lines in the TW Hya disk (\S 
3.1).  Next, we demonstrate that standard models are unable to reproduce the 
ALMA data with the typical assumptions about disk temperature profiles and the 
spatial variation of the disk velocity field projected on the sky (\S 3.2).  
Then, we present three options for simple modifications to those assumptions 
that are able to reconcile the data and models (\S 3.3): a large temperature 
increase inside the TW Hya disk cavity ($r \le 4$\,AU; \S 3.3.1); a deviation 
from a Keplerian velocity field that produces relatively enhanced rotation 
rates at small disk radii (\S 3.3.2); or a warp that produces a relatively 
higher line-of-sight inclination angle in the inner disk (\S 3.3.3).  We focus 
on a qualitative illustration of how the data reveal novel spatio-kinematic 
information on gas at small radii in the TW Hya disk, rather than employing a 
quantitative model-optimization (detailed fitting) approach: the latter will be 
more useful in the near future, when ALMA provides TW Hya data with further 
improvements in dynamic range and angular resolution (as well as sensitivity).

\subsection{Modeling Overview}

We defined a simple, azimuthal- and mirror-symmetric parameterized model for 
the disk gas densities on a cylindrical grid with coordinates ($r$, $\phi$, 
$z$), such that
\begin{equation}
\rho(r,z) = \frac{\Sigma}{\sqrt{2\pi} H_p} \exp \left[ -\frac{1}{2}\left(\frac{z}{H_p}\right)^2 \right], 
\end{equation}
where $\Sigma$ and $H_p$ are the radial surface density and pressure scale 
height profiles, respectively.  For the former, we adopted the similarity 
solution for a simple viscous accretion disk 
\citep[][see also Andrews et al.~2009, 2010]{lynden-bell74,hartmann98},
\begin{equation} 
\Sigma(r) = \Sigma_c \left(\frac{r}{r_c}\right)^{-\gamma} \exp\left[-\left(\frac{r}{r_c}\right)^{2-\gamma}\right],
\end{equation}
where $\gamma$ is the gradient parameter that describes the power-law radial 
variation of the disk viscosity, $r_c$ is a characteristic scaling radius, and 
$\Sigma_c = e \cdot \Sigma(r_c)$.  Since we considered only emission from the 
CO molecule, the models were parametrically defined in terms of the CO column 
density,
\begin{equation}
N_{\rm co}(r) = \frac{X_{\rm co}}{\mu m_H} \,\, \Sigma(r),
\end{equation}
where $X_{\rm co}$ is the CO abundance fraction in the gas, $\mu = 2.37$ is the 
mean molecular weight of the gas, and $m_H$ is the mass of a hydrogen atom.  
The scale height profile was set by the balance of gravity and thermal pressure 
in the disk, such that
\begin{equation}
H_p(r) = \frac{c_s}{\Omega} = \left[ \frac{k_b \, r^3 T(r)}{G M_{\ast} \mu m_H} \right]^{1/2},
\end{equation}
where $c_s$ is the local sound speed, $\Omega$ is the Keplerian angular 
velocity, and $k_b$ is the Boltzmann constant.  In this formulation, we 
implicitly assumed that the disk is vertically isothermal (at each radius), 
with a radial temperature profile $T(r) = T_{10} (r/{\rm 10 \, AU})^{-q}$, 
where $T_{10} = T({\rm 10 \, AU})$.  The disk kinematics are determined by a 
velocity field in the model plane, $\vec{v} = (v_r, v_{\phi}, v_z)$.  In the 
models considered here, $v_r = v_z = 0$, and the azimuthal component of 
$\vec{v}$ is described by Keplerian rotation,
\begin{equation} 
v_{\phi}(r) = v_k = \sqrt{\frac{GM_{\ast}}{r}},  
\end{equation}
assuming implicitly that the disk mass is negligible compared to the stellar 
mass.  

The structure model described in Equations (1)-(5) is fully characterized by a 
set of 6 parameters, \{$N_c$, $r_c$, $\gamma$, $T_{10}$, $q$, 
$M_{\ast}$\}.\footnote{$N_c$ is defined as the CO column density at the 
characteristic radius, $r_c$; see Equation (3).}  An additional 3 parameters 
are required to convert this structure to a synthetic CO emission line model 
projected into the sky plane: a turbulent velocity dispersion ($\xi$), and a 
disk viewing geometry specified by the inclination ($i$) and major axis 
position angle (PA).  For a given set of 9 parameters, we used the 
three-dimensional non-local thermodynamic equilibrium (non-LTE) molecular 
excitation and radiative transfer code {\tt LIME} \citep[v1.02;][]{brinch10} 
and the data from the LAMBDA database \citep{scholier05} to calculate level 
populations and generate high-resolution synthetic spectral cubes for the CO 
$J$=2$-$1 and $J$=3$-$2 emission lines.  Those cubes were Fourier transformed 
and re-sampled onto the spectral and spatial frequency grids employed in the 
ALMA observations.  The resulting model spectral visibilities were then passed 
through the same {\tt CASA} imaging processes that were performed on the data 
(see \S 2).

\subsection{The Failure of Standard Models}

\begin{figure}[t!]
\epsscale{1.00}
\plotone{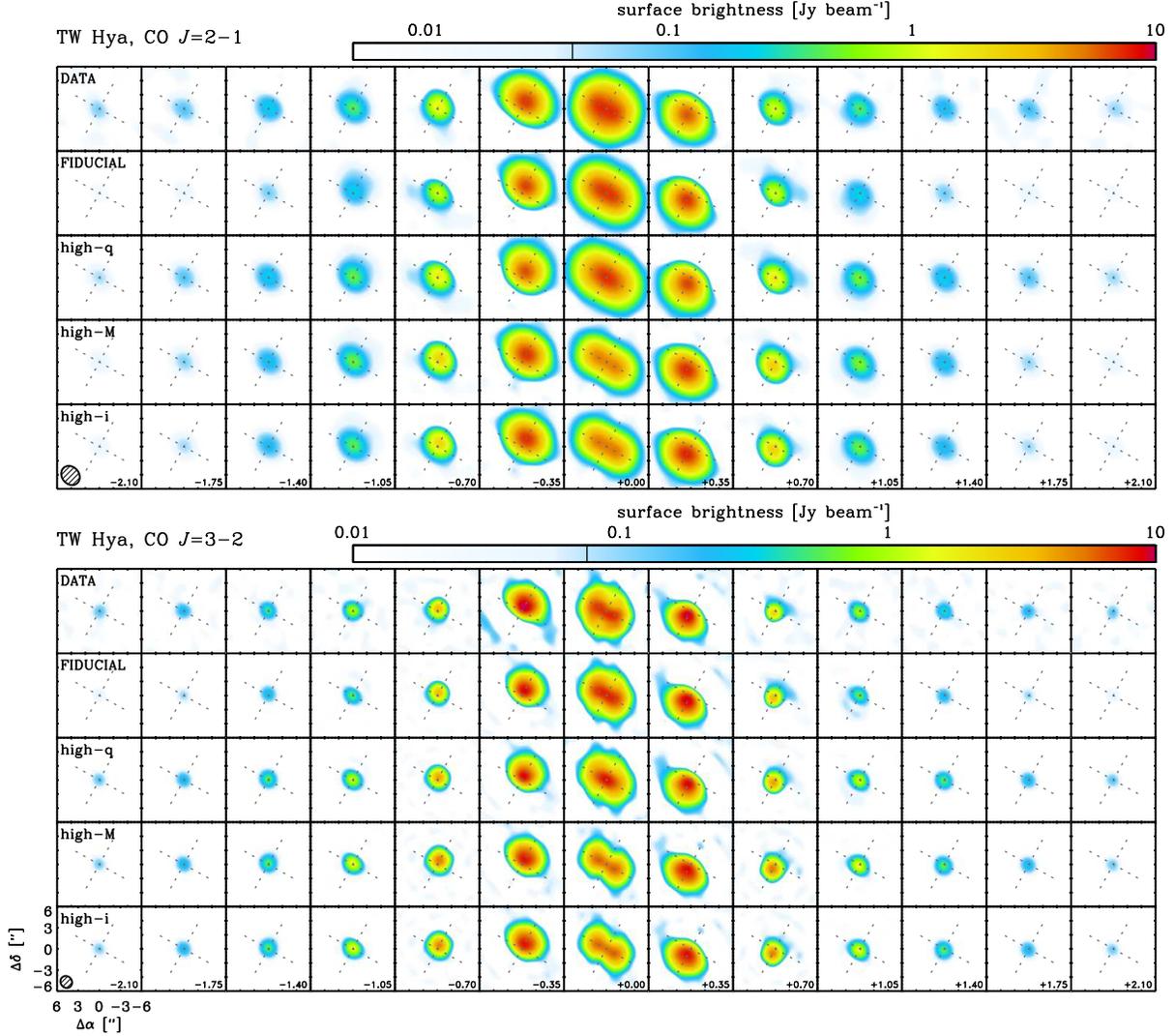}
\figcaption{Comparisons of the observed CO $J$=2$-$1 and $J$=3$-$2 channel maps 
({\it top rows}) with the fiducial model and the three other standard models 
described in \S 3.2 (see Table \ref{tab:params1} for parameter values).  The 
observed and model channel maps were synthesized on the same binned 0.35\,km 
s$^{-1}$ velocity gridding for ease of comparison.  The map sizes, intensity 
scale bars, LSR velocity labels, and synthesized beam dimensions are the same 
as in Figures \ref{fig:co21} and \ref{fig:co32}.  \label{fig:compare1}}
\end{figure}

We first constructed a ``fiducial" model using the standard parameterization 
described in \S 3.1, motivated by the results of our previous analyses of CO 
emission from the TW Hya disk \citep{qi04,qi06,hughes08,hughes11,andrews12}.  
Because of the time-consuming nature of the {\tt LIME} calculations for any 
given model, we adopted a simplistic exploration of parameter-space that 
utilized the modified Levenberg-Marquardt minimization algorithm {\tt MPFIT} 
\citep{markwardt09} to help find parameters that produced a low $\chi^2$ in the 
spectral visibilities for both CO transitions.  In that process, we forced the 
basic parameters that describe the disk densities \{$N_c$, $\gamma$, $r_c$\} 
and turbulent velocity width \{$\xi$\} to be the same for both CO transitions, 
but allowed different values for the temperature profile parameters \{$T_{\rm 
10}$, $q$\} for the two emission lines.  The fiducial stellar mass was fixed to 
$M_{\ast} = 0.8$\,M$_{\odot}$ \citep[e.g.,][]{wichmann98,hughes11}, and the 
major axis position angle was set to 151\degr\ (east of north); the disk 
inclination $\{i\}$ was considered a free parameter (albeit the same for both 
CO transitions).  Guided by the minimization results, the model parameters were 
slightly adjusted manually to best reproduce the observed channel maps (at 
native resolution).  The fiducial model parameters we derived are provided in 
Table \ref{tab:params1} \citep[not surprisingly, they are similar to those for 
the Model sA constructed by][]{andrews12}.  The model channel maps are compared 
directly with the data in Figure \ref{fig:compare1}, imaged on the binned 
velocity grid with the same channel-spacing of 0.35\,km s$^{-1}$ for both CO 
transitions, for clarity.  

The fiducial model is a good match to the ALMA data near the systemic velocity, 
but notably fails to reproduce the emission line wings that are observed for 
both CO transitions.  Those wings are observed at velocities at 2.1\,km 
s$^{-1}$ (and beyond, albeit at lower statistical significance) from the line 
core, corresponding to intrinsic disk velocities in excess of 20\,km s$^{-1}$ 
considering the inferred disk inclination of $\sim$6\degr.  For this stellar 
mass and assuming Keplerian orbits, those velocities are only reached at disk 
radii $<$2\,AU.  Since these CO lines have such high optical depths, their 
intensities scale with the product of the gas temperature and the projected 
emitting area \citep{beckwith93,dutrey94}.  With such a small area available to 
produce emission at these line wing velocities, this discrepancy between the 
fiducial model and the data is perhaps not so surprising.  Broadly speaking, 
there are two approaches that could help reproduce the observed CO emission 
line wings: the first deals with adjusting the physical conditions of the model 
at small radii, and the second relies on modifying the model parameters that 
control the projected velocity field.  

For the first (structure) approach, a representative model needs to incorporate 
higher gas temperatures at small disk radii (because the optical depth is so 
high, adjustments to the CO column densities have little effect on the line 
wing emission strengths).  Sticking with the standard prescription for modeling 
this kind of data (described in \S 3.1), that requirement effectively amounts 
to adopting a steeper temperature profile.  We constructed such a ``high-$q$" 
model with a larger radial temperature gradient ($q = 0.65$); parameter values 
are listed in Table \ref{tab:params1}, and synthetic channel maps are compared 
directly with the data and fiducial model in Figure \ref{fig:compare1}.  By 
increasing the gas temperatures inside $\sim$5\,AU by roughly a factor of 3, 
this kind of model is able to reproduce the observed intensities in the 
emission line wings.  However, the required tempertaure profile is too steep to 
properly account for the emission morphology generated at relatively large disk 
radii, and therefore ends up being a very poor match for the CO emission near 
the systemic velocity.  

The alternative (kinematic) approach relies on increasing the projected area of 
the high-velocity disk region responsible for producing the CO line wings.  Put 
simply, the observations demand an increase in the projected velocity field 
($v_{\rm obs} = v_{\phi} \sin{i}$).  This can be achieved either by scaling up 
the intrinsic disk velocities ($v_{\phi} = v_k \propto \sqrt{M_{\ast}}$) and/or 
the projected line of sight ($\sin{i}$).  To demonstrate that, we constructed 
two additional models using the standard prescription outlined in \S 3.1.  
First, a ``high-$M_{\ast}$" model employed a larger stellar mass ($M_{\ast} 
= 1.5$\,M$_{\odot}$) to speed up the Keplerian orbital velocities ($v_{\phi}$), 
thereby increasing the extent of the high-velocity emitting area.  And second, 
a similar effect was achieved for a ``high-$i$" model that adopts a more 
inclined viewing geometry ($i = 8\degr$).  Both of these models increase the 
projected line of sight velocities by a factor of $\sim$1.4, which corresponds 
to a factor of $\sim$3.5-4 increase in the emitting area that produces the 
high-velocity line wings.  The parameters for these models are listed in Table 
\ref{tab:params1}.  Their corresponding synthetic channel maps are compared 
with the data and fiducial model in Figure \ref{fig:compare1}.  As expected, 
the high-$M_{\ast}$ and high-$i$ models successfully reproduce the observed CO 
line wings: however, they consequently do a very poor job accounting for the 
morphology near the systemic velocity.

In practice, the high-$q$, high-$M_{\ast}$, and high-$i$ models that follow the 
standard formalism outlined in \S 3.1 all fail to reproduce the observed CO 
emission line morphologies {\it in the same way}.\footnote{From a statistical 
perspective, these kinds of model are formally much poorer matches to the data 
than the fiducial model: their discrepancies with the data are primarily in the 
spectral channels that contain most of the emission signal, and therefore most 
of the weight (contribution to $\chi^2$) in any quantitative assessment of fit 
quality.}  No reasonable combinations of $q$, $M_{\ast}$, and $i$ can 
simultaneously reproduce both the CO line wings and the emission morphology 
near the systemic velocity.  Uniformly scaling up the temperature profile 
gradient or projected velocity field at all disk radii cannot explain the 
patterns of the ALMA data.

\subsection{Alternative Models}

The standard modeling formalism explored in \S 3.2 is unable to explain the 
observed spatio-kinematic emission morphologies of the CO emission lines 
generated by the TW Hya disk.  However, that modeling effort demonstrated 
clearly the ``sense" of this failure.  To produce the emission line wings, we 
need a model with either much higher temperatures or increased projected 
velocities in the inner disk.  However, the observed emission morphology near 
the systemic velocity was explained best with a fiducial temperature structure 
and projected velocity field.  This implies that we require a deviation from 
the standard structure and kinematics model that {\it varies with radius} in 
the disk.  As before, these modifications to $T(r)$ or $v_{\rm obs}(r) = 
v_{\phi} \sin{i}$ can be manifested in a deviation to the form of the 
temperature profile, the intrinsic disk velocity field, $v_{\phi}(r)$, and/or 
the disk viewing angle, $i(r)$, so long as those functional forms {\it 
decrease} with $r$.  Three such alternatives were explored here, using simple 
parametric adjustments to the standard model prescription.

\subsubsection{A Hot Inner Disk}

\begin{figure}[t!]
\epsscale{1.00}
\plotone{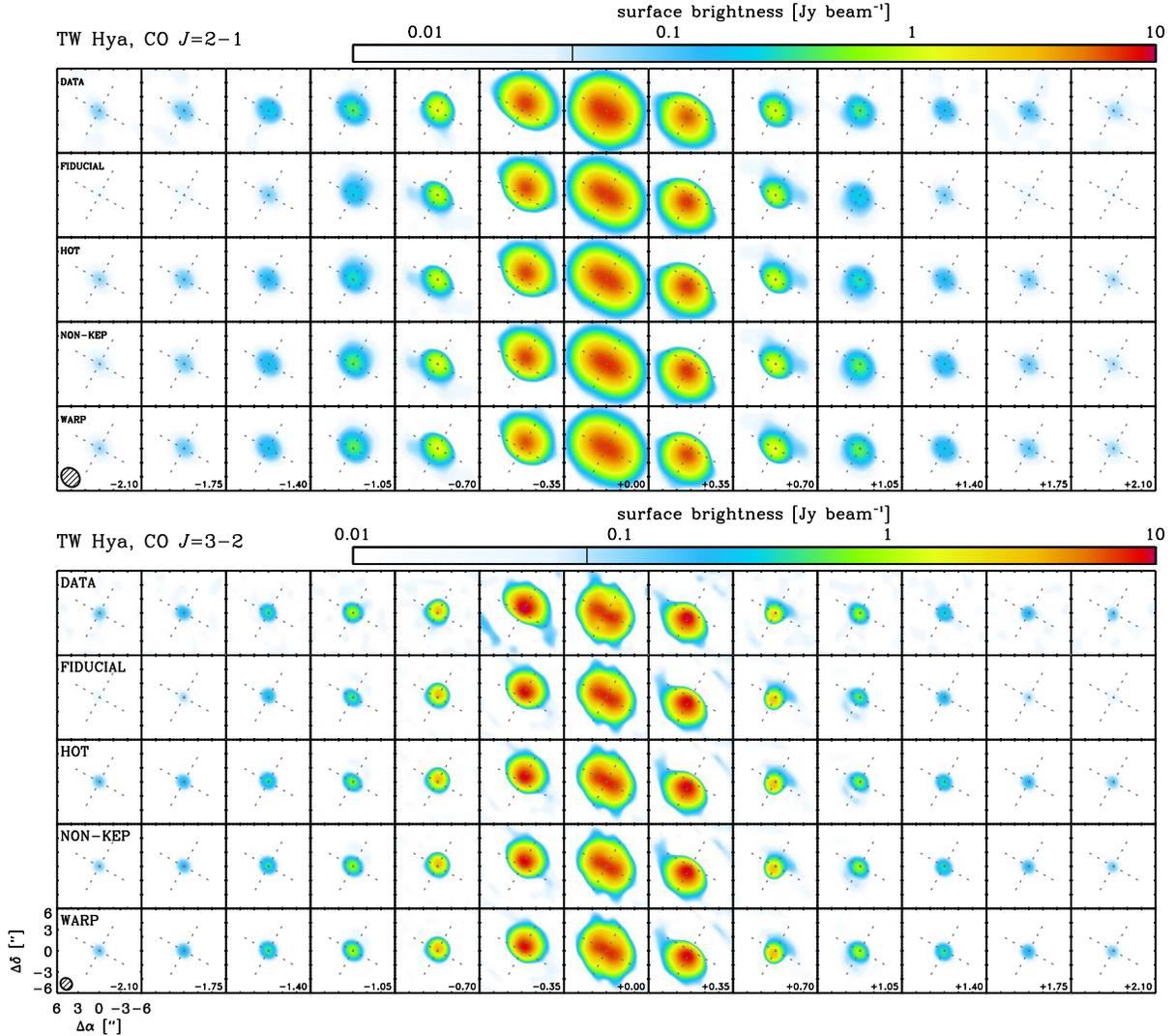}
\figcaption{A channel map comparison of the data, fiducial model, and the 
alternative models described in \S 3.3 (similar to Figure \ref{fig:compare1}).  
Model parameters are listed in Table \ref{tab:params1}.  \label{fig:compare2}}
\end{figure}

The first approach we adopted for these alternative models was to implement a 
structural modification to the radial temperature profile.  As was shown in \S 
3.2, models with a hot inner disk (steeper gradient $q$) are able to reproduce 
the observed CO line wing intensities, but the outer disk needs to have a more 
shallow temperature profile to account for the emission morphology near the 
line core.  In principle, these features could be reconciled by permitting the 
gradient parameter $q$ to vary with radius.  However, such a modification is 
not so intuitive; instead, we elected to simply permit a multiplicative scaling 
of the temperatures inside a given radius, such that
\begin{equation}
T(r) = (\delta T) \cdot T_{10} \left(r/{\rm 10 AU}\right)^{-q}
\end{equation}
where $\delta T$ is a free (constant) parameter in the inner disk, but 
$\delta T = 1$ when $r > r_{\rm hot}$.  In effect, this model adjustment 
preserves the fiducial outer disk temperature profile, but artificially scales 
up the temperatures at small disk radii to produce the CO line wings.  Based on 
this modified framework, we developed a representative model for the CO 
emission following the procedure described in \S 3.2.  The resulting parameters 
for this ``hot" model are listed in Table \ref{tab:params1}.  Synthetic 
observations of the model are compared with the data as channel maps in Figure 
\ref{fig:compare2}, as well as in position-velocity diagrams (constructed from 
a slice along the disk major axis) in Figure \ref{fig:PV1}.  

We find a good match to the CO data by modifying the fiducial model described 
in \S 3.2 with a scaling factor $\delta T = 3$ inside a radius $r_{\rm hot} = 
4$\,AU.  This choice for $r_{\rm hot}$ was not arbitrary; it was selected 
because the TW Hya disk is known to have substantially diminished dust optical 
depths inside that radius \citep{calvet02,hughes07,arnold12}.  Notably, the 
implicit assumption for this scenario is that a substantial gas reservoir is 
still present inside the dust cavity.

\subsubsection{A Non-Keplerian Velocity Field}

\begin{figure}[t!]
\epsscale{0.90}
\plotone{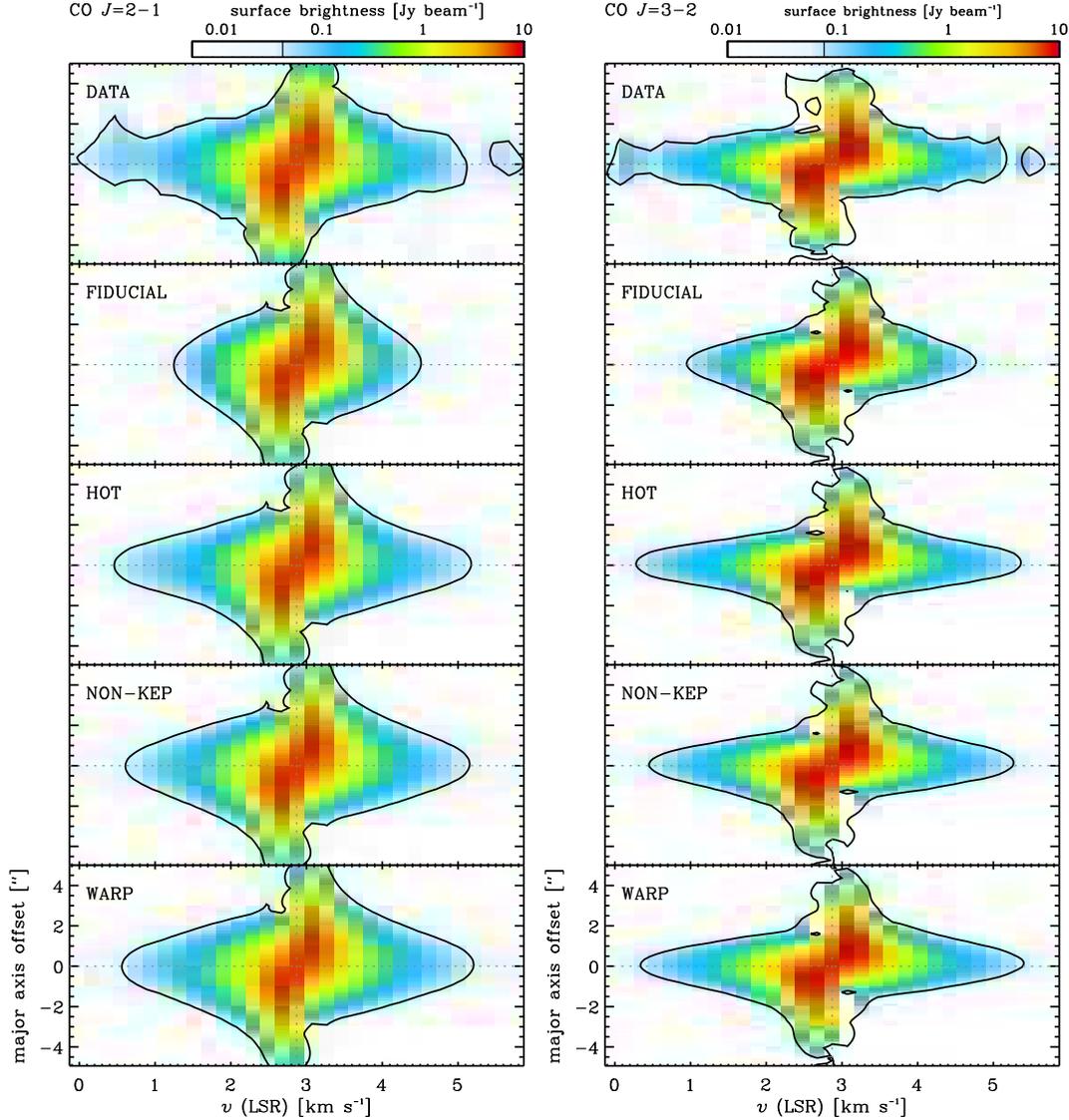}
\figcaption{CO position-velocity diagrams constructed from a cut along the disk 
major axis (P.A. = 151\degr) for the data and the fiducial, hot, non-Keplerian, 
and warped models described in \S 3.3.  For reference to gauge the extent of 
the line wings, the 5\,$\sigma$ brightness level is marked with a contour.  
\label{fig:PV1}}
\end{figure}

The second alternative we considered was a kinematic modification to the 
intrinsic velocity field of the gas in the inner disk.  Again, as demonstrated 
in \S 3.2, the outer disk velocities can not also be scaled up without 
producing large deviations in the observed emission morphology near the 
systemic velocity.  To accomodate both requirements, we introduced a simple 
scaling factor for the disk velocity field, $v_{\phi}(r) = f v_k$ (see Equation 
5).  The scaling factor $f$ describes the model deviation from Keplerian orbits 
and is permitted to vary with radius like a power-law in the inner disk, 
\begin{equation}
f(r) = \left\{ \begin{array}{ll}
                (r/r_b)^{-x} & \mbox{if $r \le r_b$} \\
                1            & \mbox{if $r > r_b$}
               \end{array}
       \right. 
\end{equation}
where \{$r_b$, $x$\} are free parameters that describe a break radius and the 
gradient of the scaling function, respectively.  In the standard models 
described in \S 3.1 and 3.2, $f = 1$ at all disk radii (effectively, $x = 0$ in 
those cases).  Adding in this parametric non-Keplerian scaling factor as a 
slight modification to the standard modeling framework, we constructed 
representative models for the CO emission lines using the procedure outlined 
above.  The resulting parameter values are listed in Table \ref{tab:params1}; 
the model is compared with the ALMA data in Figures \ref{fig:compare2} and 
\ref{fig:PV1}.

Adopting $r_b \approx 60$\,AU and $x = 0.15$ to describe $f$, we were able to 
reproduce well the observed CO line wings without sacrificing the fit quality 
of the emission morphology near the systemic velocity.  The super-Keplerian 
rotation in the inner disk means that the CO emitting area at the observed 
velocities of the line wings is substantially enhanced relative to the fiducial 
model (by a factor of $\sim$5), and therefore able to generate high-velocity 
emission at roughly the observed levels.  The \{$r_b$, $x$\} values adopted 
here are not unique (nor optimized); a smaller break radius and larger gradient 
(or vice versa) could probably accomodate the data with a similar fit quality.  
However, these values and this {\it kind} of model modification are sufficient 
to demonstrate the key point: a non-Keplerian deviation in the disk velocity 
field which decreases with radius successfully accounts for the observed 
spatio-kinematic CO structure emitted by the TW Hya disk.

\subsubsection{A Disk Warp}

We explored another kinematic alternative model that incorporates a warp in the 
vertical structure of the gas disk.  Rather than raise the intrinsic orbital 
velocity of the gas ($v_{\phi}$), this model increases the observed velocity by 
changing the line of sight ($\sin{i}$) as a function of radius.  In this way, a 
warp -- or a disk inclination that {\it increases} toward the central star -- 
raises the projected velocities in the inner disk, effectively increasing the 
emitting area at a given observed velocity.  If the inclination is a decreasing 
function of $r$, this model can be designed to have minimal impact on the 
kinematic pattern for the outer disk, promoting a favorable emission morphology 
near the systemic velocity.  To construct a warped disk model, we introduced a 
power-law profile for the observed inclination angles, 
\begin{equation}
i(r) = i_0 \left(\frac{r}{r_0}\right)^{-y}, 
\end{equation}
aligned with the disk major axis (assuming a vertically thin disk; see the 
Appendix), where \{$i_0$, $y$\} are parameters that describe the inclination 
normalization (at $r_0$) and its radial gradient, respectively.  For a given 
set of structure parameters, the model densities and temperatures are rotated 
to this modified viewing geometry.  The velocity field is tilted using a 
quaternion transformation, as described in more detail in the Appendix.  
Incorporating this parametric formulation for a warped disk viewing geometry, 
we constructed models for the TW Hya CO emission lines as before.  The 
resulting parameter values for a representative warped model are included in 
Table \ref{tab:params1}; synthetic observations of the models are compared with 
the data in Figures \ref{fig:compare2} and \ref{fig:PV1}.  

Models that include a modest warp, with $i_0 \approx 8\degr$ at $r_0 = 5$\,AU 
and a gradient $y = 0.15$, provide an excellent match with the ALMA data for 
both CO emission lines, reproducing well the observed line wings and emission 
morphology near the systemic velocity.  Like the non-Keplerian model described 
in \S 3.3.2, this warp formulation successfully produces high-velocity line 
emission by increasing the projected CO emitting area at the observed wing 
velocities by a factor of $\sim$5 (by design).  And also like that case, the 
warp model parameters we determine or even the formulation we adopt are not 
unique, nor optimized.  Instead, they are intended to conceptually verify that 
a warped disk geometry can reproduce in detail the fundamental features of the 
observed CO line emission with the addition of only a single model parameter; 
an inclination gradient, $y$.

\subsection{Additional Comments}

It is worth pointing out that the alternatives discussed in the previous 
subsections are in no way mutually exclusive: various combinations of these 
effects could be used to achieve similar results.  Moreover, there are 
additional alternatives to a simple disk model that could presumably account 
for these ALMA observations, but have not been explored in detail here.  
Perhaps most compelling among these is a model that explicitly incorporates a 
spatial variation in the broadening term of the line profile.  For example, it 
seems reasonable that a model with a turbulent line width $\xi$ that decreases 
with radius should be able to re-distribute some emission into the 
high-velocity line wings observed here.  The $\xi$ values we adopted are 
commensurate with the turbulence constraints imposed by \citet{hughes11} for 
the outer disk, made using observations with $\sim$4-5$\times$ higher spectral 
resolution than the ALMA data.  However, we cannot rule out higher $\xi$ values 
at small disk radii.  We attempted a simple modification to our fiducial model 
that permits the turbulent velocity widths to scale with the sound speed of the 
gas, using the \citet{hughes11} upper limit to normalize $\xi(r)$ in the outer 
disk: but given the low $\xi$ at large $r$ and modest radial variation of 
$c_s$, this has little effect on the CO line wings.  If a variable turbulent 
line width parameter is the origin of these wings, $\xi$ must increase more 
steeply than $c_s$ in the inner disk.  Given the limited velocity resolution of 
these ALMA data relative to the subsonic turbulent velocity dispersions in the 
TW Hya disk, we felt it was premature to push the further exploration of such 
models.  Nevertheless, it remains an interesting possibility that could be 
investigated with future observations.

\section{Discussion}

We have conducted a detailed analysis of the $^{12}$CO $J$=2$-$1 and $J$=3$-$2 
line emission from the gas-rich disk around the nearby young star TW Hya, 
making extensive use of the science verification data from the Atacama Large 
Millimeter Array (ALMA) commissioning effort.  With the substantial improvement 
in line sensitivity afforded by ALMA (even during its construction), we 
identified a novel feature in these CO spectra: the presence of relatively 
faint line wing emission extending out to high projected velocities of at least 
$\pm2.1$\,km s$^{-1}$ from the (systemic) line center, roughly twice the 
maximum velocities previously inferred for the TW Hya disk from less sensitive 
data \citep{qi04,qi06,hughes11,andrews12}.  Given the nearly face-on viewing 
geometry of the TW Hya disk, gas on Keplerian orbits that generates these line 
wings must be located within a few AU from the central star.  This kinematic 
super-resolution was employed in an exploration of the temperature and velocity 
structure of the inner disk, utilizing a non-local thermodynamic equilibrium 
molecular excitation and radiative transfer code and a standard modeling 
prescription for circumstellar disks.  We found that the typical assumptions of 
a simple, Keplerian disk with a smooth gas temperature profile were unable to 
simultaneously produce both the CO line wings and the observed line emission 
morphology near the systemic velocity.  Instead, we developed three alternative 
model possibilities that are able to qualitatively account for the ALMA data by 
making small (parametric) adjustments to the spatio-kinematic structure of the 
gas in the inner disk.  

In one such alternative, we considered a parametric scaling of the gas 
temperatures inside the known dust-depleted cavity at the disk center \citep[$r 
< 4$\,AU;][]{calvet02,hughes07} in an effort to boost the CO line wing 
intensities without substantially disturbing the emission morphology near the 
line core.  With the outer disk structure based on a fiducial model, an 
increase of the gas temperatures in the dust cavity by a factor of $\sim$3 
successfully accounts for the ALMA CO observations.  While such a temperature 
scaling might seem simplistic and artificial, it is in fact consistent with continuum radiative transfer models for the TW Hya disk: the low optical 
depths inside the disk cavity mean that stellar radiation is effectively 
unattenuated, leading to substantial heating \citep{calvet02,andrews12}.  If we 
assume that gas and dust temperatures are correlated (if not identical), then 
the factor $\delta T \approx 3$ derived here is quite reasonable.  For 
reference, the models developed by \citet{andrews12} have dust temperatures 
that increase by a factor of $\sim$4 over a narrow ($\sim$1\,AU-wide) region 
around the cavity radius (see their Figure 3).  It is interesting to consider 
the key implication of this model: the CO emission line wings detected by ALMA 
may offer an indirect (and not very practical) observational signature of a 
low-density dust cavity, but they also {\it require} that this cavity still 
hosts a significant reservoir of molecular gas.  

Of course, the presence of gas inside the TW Hya disk cavity has been inferred 
from other diagnostics: accretion indicators \citep{muzerolle00}, as well as 
emission lines of simple atoms and molecules \citep{najita10}, including CO 
\citep[e.g.,][]{rettig04}.  Based on a rotational diagram constructed from the 
infrared spectrum of CO fundamental ($v$ = 1$-$0) rovibrational emission lines, 
\citet{salyk07} used a simple slab model to estimate $T = 650$-1500\,K and 
$N_{\rm co} = 0.2$-$1.4 \times 10^{19}$\,cm$^{-2}$ for an emitting region that 
corresponds to $r \approx 0.1$-1\,AU.  Taking the mean temperature from the two 
CO lines, we find $T = 640$-1660\,K and $N_{\rm co} = 14$-$140\times 
10^{19}$\,cm$^{-2}$ for our ``hot" model in the same region -- remarkably 
similar, keeping in mind that column density estimates from optically thick 
lines are inherently ambiguous.  However, other analyses of the CO 
rovibrational spectrum suggest much lower temperatures \citep[by a factor of 
$\sim$2;][]{rettig04,pontoppidan08,salyk09}.  Moreover, our ``hot" model 
assumes enhanced temperatures over a continuous emitting area out to 4\,AU: if 
a smaller area were adopted, the temperature scaling factor $\delta T$ would 
have to be proportionately increased to compensate for the observed line wing 
intensities (notably to uncomfortable levels that start to impinge on the 
thermal dissociation energy).  Given the uncertainties involved in analyzing 
both kinds of CO data, it is not clear whether a ``hot" model derived from the 
rotational lines (as in \S 3.3.1) is necessarily in conflict (i.e., too hot) 
with the rovibrational spectrum.  In any case, it is interesting to note that 
these two complementary methods could eventually be used to obtain independent 
constraints on the gas properties inside the TW Hya dust disk cavity.  In the 
near future, ALMA can provide sensitive high angular resolution images of these 
same CO transitions that will resolve the spatio-kinematic morphology of their 
line wings.  That information will be crucial for locating the CO emitting 
area, enabling more direct comparisons between the infrared and millimeter-wave 
diagnostics.  

While the first model described above was devoted to modifications of the {\it 
structure} in the inner disk, the remaining two alternatives focused on 
adjustments to the disk {\it kinematics}.  Rather than increasing the intrinsic 
line intensities from a fixed emitting area, these other models produced the 
observed CO line wings by increasing the effective emitting area itself.  In 
one such scenario, we modified the intrinsic velocity field in the inner disk 
with a simple parametric deviation from Keplerian orbits.  We found in \S 3.3.2 
that a model which incorporated super-Keplerian rotation in the inner disk 
successfully reproduced the observed CO line emission.  Although there is 
considerable uncertainty involved, our ``non-Keplerian" model rotated 
$\sim$30\%\ faster than Keplerian speeds at $r = 10$\,AU, and $\sim$80\%\ 
faster at 1\,AU ($f \approx 1.3$ and 1.8, respectively).  If this deviation was 
restricted to much smaller disk radii only, the corresponding scaling factor 
$f$ would need to be substantially larger to account for the observed CO line 
wings.  Admittedly, of the three alternative models explored here, this 
scenario is the least physically-motivated.  While in principle a (reverse) 
pressure gradient could accelerate the disk material, most models that consider 
such gradients instead call for sub-Keplerian motions 
\citep[e.g.,][]{weidenschilling77,shu07,shu08}.  However, it is possible that 
non-azimuthal gas velocities could mimic the parameterization used here.  For 
example, a molecular outflow with high mass-loss rates, directed primarily in 
the vertical direction (a $v_z$ component) and launched from the inner disk, 
might play a role.  The TW Hya disk is thought to drive a substantial 
photoevaporative wind \citep[e.g.,][]{pascucci11}, but the molecular properties 
of that flow have not yet been explored in sufficient detail to easily compare 
with observations.  Another example might be a velocity field with radial 
streaming motion (a $v_r$ component), perhaps directed along spiral arms 
\citep[e.g.,][]{quillen06}.  No such structural features have yet been 
identified for the TW Hya disk; but there are similar disks that do exhibit 
this kind of asymmetry \citep{weinberger99,augereau99,grady01,clampin03,fukagawa04,fukagawa06,muto12}.  Though these possibilities seem exotic, they remain in 
consideration until realistic, testable models for more complex disk velocity 
fields are developed.
 
The final alternative model we developed to account for the ALMA CO data also 
relied on modifying the {\it observed} disk velocity field, $v_{\rm obs} = 
v_{\phi} \sin{i}$.  In this scenario, we maintained the standard assumption of 
Keplerian orbits ($v_{\phi} = v_k$) and instead permitted the projection factor 
$\sin{i}$ to vary with radius.  In \S 3.3.3, we developed a simple parametric 
description of a disk warp that simultaneously accounts for the CO line wings 
and the nearly face-on emission morphology near the systemic velocity.  That 
simple model has an inclination of $\sim$4\degr\ in the outer disk and 
increases only slightly to $\sim$10\degr\ at $r = 1$\,AU; the mean 
line-of-sight viewing angle averaged over the disk area is comparable to the 
$\sim$4-7\degr\ inferred previously \citep{krist00,weinberger02,qi04,hughes11}. 
Warps have been directly imaged in scattered light for the edge-on (effectively 
gas-free) debris disks around $\beta$ Pic \citep{kalas95,heap00} and AU Mic 
\citep{liu04,krist05}, and indirectly inferred from quasi-periodic photometric 
variations in optical/infrared light curves in other edge-on cases, notably KH 
15D \citep{chiang04} and AA Tau \citep{bouvier99}.  Warped viewing geometries 
have also been speculated for a few additional cases, including the disks 
around GM Aur \citep[from CO spectral imaging;][]{dutrey98,hughes09} and HD 
100546 \citep[from scattered light images;][]{quillen06b}.  

In the specific case of the TW Hya disk, \citet{roberge05} have cited potential 
evidence for a warp based on an azimuthal asymmetry in their wideband {\it 
Hubble Space Telescope} ({\it HST})/STIS coronagraphic images of optical light 
reflected off the disk surface.  They identify a roughly sinusoidal variation 
of scattered light with the disk position angle in a region 
$\sim$1.3-1.6\arcsec\ from the central star, with a peak near P.A. = 234\degr\ 
and a trough near 54\degr\ (measured E of N, and notably roughly $\pm$90\degr\ 
from the disk major axis orientation).  \citet{roberge05} speculated that this 
azimuthal modulation in the brightness profile might be the consequence of a 
varying illumination pattern in the outer disk, generated by the shadowing 
effect of a warped disk geometry at much smaller radii.  This is effectively 
the same type of model we have employed to explain the ALMA observations of the 
CO lines, so naturally we were motivated to test this possibility against the 
behavior of the \citet{roberge05} scattered light data.  To do so, we 
implemented a simplified dust structure based on the ``warp" model described in 
\S 3.3.3, assuming that the dust and gas are co-located outside a radius of 
4\,AU: to adhere to the dust modeling in the literature, we assume there is 
effectively no dust for $r < 4$\,AU \citep[e.g.,][]{calvet02,hughes07}.  We 
adopt the relative mass ratio, grain composition, and size distribution used in 
the disk atmosphere models of \citet{andrews12}.  The three-dimensional Monte 
Carlo radiative transfer code {\tt RADMC-3D}\footnote{\url{http://www.ita.uni-heidelberg.de/$\sim$dullemond/software/radmc-3d/}} (v0.30: C.~P.~Dullemond) was 
then used to generate a high-resolution synthetic image at a wavelength of 
0.8\,$\mu$m, assuming isotropic scattering.  Following \citet{roberge05}, we 
calculated a normalized intensity profile in 20\degr\ azimuthal bins in an 
annular ring from $r = 70$-88\,AU.  

\begin{figure}[t]
\epsscale{0.85}
\plotone{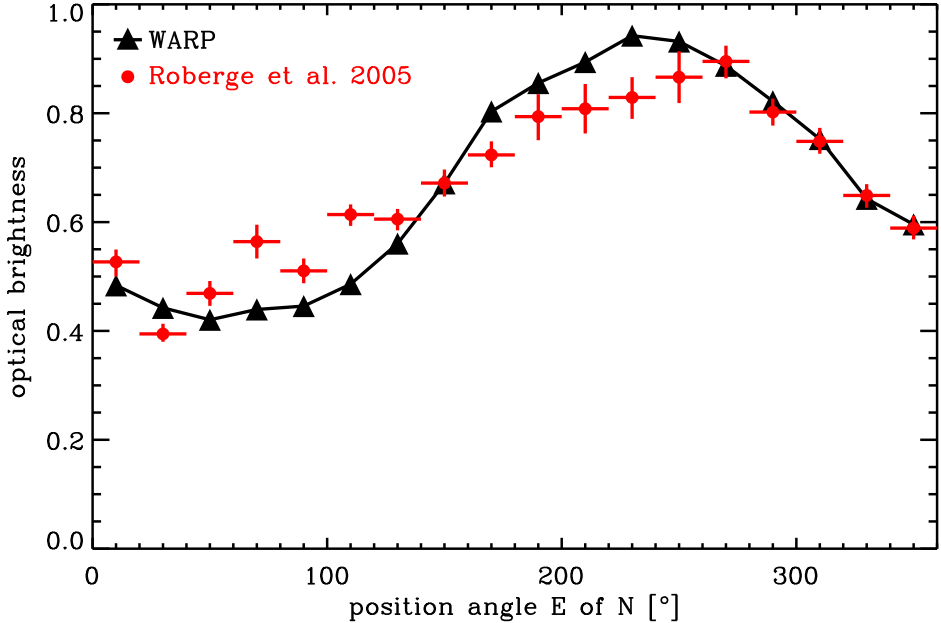}
\figcaption{A comparison of the (normalized) azimuthal profile of optical light 
scattered off the TW Hya disk surface from both {\it HST}/STIS coronagraph data 
\citep[{\it red circles};][]{roberge05} and a prediction based on our ``warp" 
model ({\it black curve}; note that this is {\it not} a fit to the scattered 
light data).  Both profiles were derived by averaging the intensities in 
20\degr\ bins from an annulus extending from $r = 70$-88\,AU.  The sinusoidal 
modulation is a natural consequence of shadowing by a warped structure in the 
inner disk (higher inclination angles at small radii), which produces a varying 
illumination pattern of the disk surface at larger radii.  \label{fig:az_scat}}
\end{figure}

The results are shown in Figure \ref{fig:az_scat} and demonstrate a striking 
resemblance to the azimuthal modulation pattern inferred from the {\it 
HST}/STIS data \citep[][see their Figure 9]{roberge05}.  Taken at face value, 
this remarkable agreement might be considered compelling independent evidence 
for a very modest warp in the TW Hya disk structure.  Again, this scattered 
light model is {\it not} a fit to the {\it HST}/STIS observations of scattering 
off dust grains, but rather the predicted behavior based on one model that is 
able to describe ALMA observations of the CO gas structure.  The sinusoidal 
pattern of the azimuthal profile is a result of material in the inner disk -- 
where the warp is enhanced -- shadowing dust at larger radii.  The phase of 
that pattern is set by the assumed warp (rotation) axis, which was arbitrarily 
aligned with the observed major axis of the disk projected on the sky 
(151\degr\ E of N).  However, in practice the phase depends only relatively 
weakly on the warp axis orientation: shifting the latter by $\pm$20\degr\ leads 
to only modest changes to the pattern phase (i.e., the data are not quite able 
to quantify the warp orientation).  The pattern amplitude depends more 
sensitively on the adopted model parameters, as it is set primarily by the 
height of the optically thick dust layer in the inner disk.  If the inner disk 
were substantially cooler, and therefore had lower scale heights, the level of 
shadowing would decrease and the amplitude of the azimuthal scattering pattern 
would be diminished.  Indeed, if we instead adopt the cooler, midplane dust 
temperatures found by \citep{andrews12} to set the dust scale heights using 
Equation 4, we find a modest ($\sim$20\%) decrease in the peak amplitudes that 
are more consistent with the scattered light data.  Perhaps more important in 
this case is the truncation of the dust distribution for $r < 4$\,AU.  If we 
were to assume no dust depletion in the disk cavity, the shadowing of the outer 
disk would be much more pronounced, with a normalized amplitude of $\sim$1 in 
the azimuthal variation of scattered light.  In essence, there is a substantial 
degeneracy between the spatial distribution of dust in the inner disk (i.e., 
the cavity size and depletion factor, as well as the local scale heights) and 
the parameters that describe the warp.  Nevertheless, for a set of simple 
assumptions grounded in the ALMA CO data \citep[and the dust cavity from the 
SED and resolved images;][]{calvet02,hughes07,andrews12}, we find a warp model 
is remarkably consistent with a key feature of the \citet{roberge05} scattered 
light data.

Warps have been postulated to originate from a wide range of phenomena, 
including gravitational instabilities \citep[e.g.,][]{sellwood10}, intense 
radiation fields \citep{armitage97}, star-disk magnetic interactions 
\citep[e.g.,][]{terquem00,flaherty10}, and dynamical perturbations from a 
stellar flyby \citep{quillen06b,nixon10}.  However, none of these seem 
particularly plausible in this case, given the known properties of TW Hya and 
its disk.  In principle, a massive planet embedded in the inner disk with an 
orbit inclined out of the disk plane could also generate the modest warp 
inferred here \citep{lubow01,marzari09}.  While this scenario has been 
extensively explored for the $\beta$ Pic debris disk 
\citep{mouillet97,dawson11}, the high gas densities in the TW Hya disk could 
lead to the substantial damping of a planet inclination -- and therefore the 
warp structure -- on a relatively short (viscous) timescale 
\citep{papaloizou95}.  However, given the large uncertainties on the disk 
viscosities and densities, as well as the range of potential companion 
properties (e.g., masses, orbits), there is still a reasonable likelihood that 
a companion could maintain the modest orbital inclination required to sustain a 
small warp over the lifetime of the TW Hya disk \citep[e.g.,][]{bitsch11}.  
Moreover, other effects related to planet-planet scattering in the system could 
produce repeated dynamical excitations of the gas disk 
\citep[e.g.,][]{thommes03}.  

We would be remiss not to point out that there is an alternative explanation 
for the \citet{roberge05} scattered light asymmetry: namely, a spatial 
variation in the scattering phase function of the grains.  This latter 
possibility might better account for the observed kinks in the {\it radial} 
scattered light profiles, which incidentally our toy model does not reproduce 
well (although perhaps could, given a considerable modeling effort beyond the 
scope of this article).  Other studies of the TW Hya scattered light disk from 
{\it HST} make no clear mention of an azimuthal asymmetry 
\citep[e.g.,][]{krist00,weinberger02}.  The original WFPC2 discovery paper by 
\citet{krist00} does note that the ``right" (we assume west) side of the disk 
appears brighter (consistent with the Roberge et al.~asymmetry), but they 
suggest that the discrepancy could just as easily be explained by an 
instrumental or calibration artifact.  We should also note that 
\citet{pontoppidan08} used infrared spectro-astrometry measurements of the CO 
fundamental rovibrational lines to determine an inclination of only 
$\sim$4-5\degr\ in the inner $\sim$0.1-0.5\,AU of the TW Hya disk.  The data 
presented here are not directly sensitive to such small radial scales, but the 
\citet{pontoppidan08} claim of a warped {\it outer} disk (where $i$ is an 
increasing function of $r$) is clearly not consistent with the ALMA data.  
These inferred variations in inclination all could be compatible if the disk 
tilt is not monotonic with radius, but instead has some maximum inclination at 
a few AU.  In any case, a simple warp model is a compelling potential 
explanation for both the ALMA CO spectral images and the {\it HST}/STIS 
scattered light data, but it is not necessarily the correct (or unique) 
solution.

In principle, all three of the alternative models that successfully explain 
the ALMA CO data could ultimately be produced by dynamical interactions between 
the disk and a companion located a few AU from the central star.  Nevertheless, 
despite the known dust depletion at small radii or potential evidence for a 
warp in this disk, there is not yet any firm evidence for a close companion to 
TW Hya.  The most definitive high-resolution contrast limits for this system 
rule out companion masses $>$0.014\,M$_{\odot}$ at 2-4\,AU separations 
\citep{evans12}, suggesting that any perturber would need to be substellar or 
planetary in nature.  A previous claim for a ``hot Jupiter" \citep{setiawan08} 
has been refuted \citep{huelamo08,rucinski08}, and, although there is a 
suggestive asymmetry in new mid-infrared data that is consistent with a faint 
point source near the edge of the disk cavity, its interpretation remains 
uncertain \citep{arnold12}.  Along with deeper companion searches in the near 
future, we expect that some of the techniques presented here could be used to 
help provide new, empirical constraints on how dynamical interactions affect 
the kinematic and structural properties of the TW Hya gas disk and others like 
it.

\section{Summary}

We have studied the spatio-kinematic properties of the CO gas in the disk 
around the nearby young star TW Hya, making use of the ALMA science 
verification datasets provided during the facility construction.  While these 
spectral images of the CO $J$=2$-$1 and $J$=3$-$2 emission lines from ALMA have 
only modest angular and spectral resolutions (and dynamic range), they provide 
roughly an order of magnitude improvement in sensitivity when compared to 
previous datasets \citep{qi04,qi06,hughes11,andrews12}.  The key results of our 
analysis include:
\begin{enumerate}
\item We find clear evidence for substantial CO emission in high-velocity line 
wings, extending out to at least $\pm$2.1\,km s$^{-1}$ from the systemic 
velocity.  Assuming Keplerian rotation for a nearly face-on disk viewing 
geometry, the projected velocities of these wings correspond to intrinsic disk 
velocities $>$20\,km s$^{-1}$, indicating that they trace molecular gas at disk 
radii $\le$ 2-5\,AU.  
\item Simple models that follow the standard prescription for analysing 
optically thick molecular lines are unable to simultaneously reproduce both the 
CO line wings and the spatially resolved emission morphology near the line 
cores.  Instead, the ALMA data suggest that the inner regions of the TW Hya 
disk must have either much higher gas temperatures and/or a faster projected 
velocity field compared to what would be expected from emission in the outer 
disk.
\item Guided by those findings, we developed three alternative models that 
reproduce the ALMA data well by including small parametric modifications to the 
standard modeling prescription.  One makes a substantial adjustment to the disk 
structure by scaling up the gas temperatures inside the known dust-depleted 
cavity in the TW Hya disk ($r < 4$\,AU).  Two others include alterations to 
parameters describing the {\it observed} disk velocity field: the first by 
permitting the intrinsic velocities in the inner disk to increase to 
super-Keplerian rates, and the second by allowing the line-of-sight viewing 
angle of the disk to increase close to the central star.  The latter model of a 
disk warp finds some independent support in its ability to account for the 
azimuthal asymmetry noted in an optical scattered light image 
\citep{roberge05}.
\item Although no one model is explicitly favored by the current suite of ALMA 
data, they all include one implicit requirement: that there be a significant 
reservoir of molecular gas inside the $\sim$4\,AU-radius region in the TW Hya 
disk known to have diminished dust optical depths.
\end{enumerate}

Perhaps most important, this study illustrates the impending capability of ALMA 
for providing a substantial advance in our understanding of molecular gas in 
protoplanetary disks.  In principle, the powerful sensitivity of ALMA and the 
kind of kinematic super-resolution highlighted here will offer novel access to 
the characteristics of gaseous material in the innermost regions of these 
disks.  In the near future, those capabilities will be supplemented with 
additional detail, as ALMA pushes toward significant improvements in both 
dynamic range and angular resolution as well.

\acknowledgments We are grateful to Rebekah Dawson, Diego Mu{\~n}oz, Christian 
Brinch, and Joanna Brown for their helpful suggestions, as well as the ALMA 
commissioning and science verification team for kindly providing detailed 
instructions that guided the calibration of these data.  A.M.H.~is supported by 
a fellowship from the Miller Institute for Basic Research in Science.  The 
Atacama Large Millimeter/submillimeter Array (ALMA), an international astronomy 
facility, is a partnership of Europe, North America and East Asia in 
cooperation with the Republic of Chile.  This article makes use of the 
following ALMA Science Verification data: ADS/JAO.ALMA\#2011.0.00001.SV.

\appendix

\section{Notes on Three-Dimensional Rotations with Quaternions}

In the general case, the rotation of a three-dimensional vector quantity can 
require some unwieldy algebraic manipulation.  A much more elegant and 
straightforward technique to accomplish the same result relies on the use of 
the quaternion algebra developed by W.~R.~Hamilton in the 1840s.\footnote{The 
original series of articles -- entitled {\it On Quaternions, Or On a New System 
of Imaginaries in Algebra} and published in installments from 1844-1850 -- is a 
recommended resource; it has been transcribed and edited by David R.~Wilkins 
and graciously posted online: see \url{http://www.emis.ams.org/classics/Hamilton/OnQuat.pdf}.}  Although quaternions have been used in many aspects of physics 
(notably in the Dirac formulation of quantum mechanics) and are commonplace in 
fields like aerospace engineering and computer graphics, we have only rarely 
come across them in the astrophysics literature.  If the warp structure 
postulated in \S 3.3.3 turns out to be a common disk feature, it might be 
useful to provide a brief tutorial on the use of quaternions in 
three-dimensional vector rotations.

Quaternions are a generalization of complex numbers.  Any quaternion ${\bf q}$ 
is represented by the combination of a scalar $s$ and a vector $\vec{v} = (v_x, 
v_y, v_z)$, such that
\begin{equation}
{\bf q} = [s, \vec{v}] = [s, v_x, v_y, v_z].
\end{equation}
Quaternion algebra has non-commutative multiplication
\begin{equation}
{\bf p} \star {\bf q} = [r, \vec{u}] \star [s, \vec{v}] = [r \, s - \vec{u} \cdot \vec{v}, \,\, r \, \vec{v} + s \, \vec{u} + \vec{u} \times \vec{v} ],
\end{equation}
an inner product, ${\bf p} \cdot {\bf q} = [r, \vec{u}] \cdot [s, \vec{v}] = r 
\, s + \vec{u}\cdot\vec{v}$, and conjugation, ${\bf \bar{q}} = [s, -\vec{v}] = 
[s, -v_x, -v_y, -v_z]$.
For the specific task of applying a rotation, the definitions of an angle 
$\theta$ and a rotation axis $\vec{a}$ can be compactly encoded in a 
quaternion, where
\begin{equation}
{\bf q} = [\cos{(\theta/2)}, \,\, \vec{a} \sin{(\theta/2)}].
\end{equation}
Any vector $\vec{v}$ can be properly mapped into its rotated counterpart 
$\vec{w}$ by generating corresponding quaternions with zero scalar components, 
${\bf v}_0 = [0, \vec{v}]$ and ${\bf w}_0 = [0, \vec{w}]$, and then applying 
a simple quaternion algebra manipulation,
\begin{equation}
{\bf w}_0 = {\bf q} \star {\bf v}_0 \star {\bf \bar{q}}.
\end{equation}

\paragraph{A Worked Example:}

To demonstrate more concretely how the formalism described above can be used in 
practice, we describe the methodology used in \S 3.3.3 to apply a warp rotation 
to the model disk described in \S 3.1.  For simplicity, we assume that the 
rotation axis corresponds with the Cartesian $x$-axis, so that $\vec{a} = (1, 
0, 0)$ by definition.  In our models, this axis is by default aligned with the 
observed major axis position angle, oriented 151\degr\ E of N projected on the 
sky.  We considered a rotation by an angle $i(r)$, which depended on the 
distance from the model center, $r = \sqrt{x^2+y^2}$ (see \S 3.3.3; Equation 
8).  In that context, consider a parcel of gas located at $\vec{v} = (x_p, y_p, 
z_p)$ where $x_p = z_p = 0$ and $y_p = 1$\,AU (i.e., 1\,AU along the $y$-axis, 
perpendicular to the warp rotation axis, and in the disk midplane).  The 
quaternion ${\bf v}_0 = [0, \vec{v}]$ describes the parcel position.  Likewise, 
the warp -- a radially varying rotation -- is also encoded in a quaternion,
${\bf q} = [\cos{(i/2)}, \sin{(i/2)}, 0, 0]$.

\begin{figure}[t]
\epsscale{1.0}
\plotone{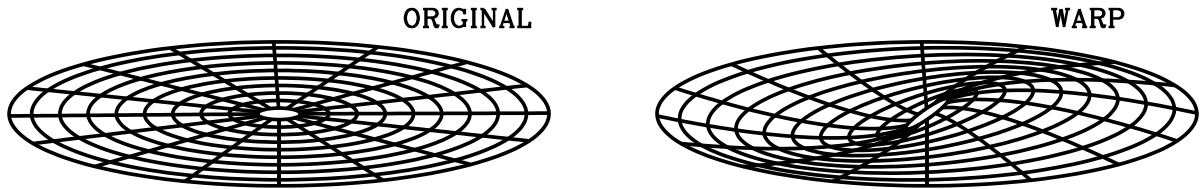}
\figcaption{Schematic of a flat model grid ({\it left}) in the disk plane ($z = 
0$) and its warped counterpart after a quaternion rotation ({\it right}).  
Given the modest rotation postulated in \S 3.3.3, the grids are shown with an 
extreme aspect ratio of 1:1:20 to highlight the subtle warp about the $x$-axis. 
\label{fig:surf}}
\end{figure}

The warp can be applied following Equation A4, and will result in a rotated 
position for the initial gas parcel that corresponds to the vector component of 
the quaternion
\begin{eqnarray}
{\bf w}_0 &=& [\cos{(i/2)}, \sin{(i/2)}, 0, 0]  \star  [0, 0, y_p, 0] \star  [\cos{(i/2)}, -\sin{(i/2)}, 0, 0] \nonumber \\
&=& [0, 0, y_p(\cos^2{(i/2)}- \sin^2{(i/2)}), \, 2 y_p \cos{(i/2)}\sin{(i/2)} ] \nonumber \\
&=& [0, 0, y_p \cos{i}, y_p \sin{i}],
\end{eqnarray}
where the final step employed elementary trigonometric identities to recover 
the simple result expected for a cell originally located orthogonal to the 
rotation axis.  Using the parameters in Table \ref{tab:params1} to reconstruct 
the angle $i$, the updated location of the gas parcel after rotation is 
described by ${\bf w}_0 = [0, 0, 0.98 \, {\rm AU}, 0.18 \, {\rm AU}]$; that is, 
slightly closer to the rotation axis and now lying at some modest height ($z = 
0.18$\,AU) above the midplane.  Repeating this approach for all model locations 
initially in the midplane ($z_p = 0$) produces the warped structure sketched in 
Figure \ref{fig:surf}.

Of course, the algebra involved in reproducing Equation A5 is substantially 
more complex for a more general location $\vec{v}$, validating the simplicity 
in the quaternion approach encapsulated in Equation A4.  For interested users, 
there are established and well-documented quaternion manipulation packages 
available online for both {\tt 
IDL}\footnote{\url{http://cow.physics.wisc.edu/$\sim$craigm/idl/math.html}} and 
{\tt python}\footnote{\url{http://cxc.harvard.edu/mta/ASPECT/tool$_-$doc/pydocs/Quaternion.html}} (among other common coding packages).

\begin{deluxetable}{lcccc}
\tablecolumns{5}
\tablewidth{0pc}
\tablecaption{Synthesized Channel Map Properties \label{tab:data}}
\tablehead{
\colhead{line} & \colhead{$\Delta v$} & \colhead{noise (1\,$\sigma$)} & \colhead{peak $I_{\nu}$} & \colhead{$F_{\nu}$} \\ 
\colhead{} & \colhead{[km s$^{-1}$]} & \colhead{[Jy beam$^{-1}$]} & \colhead{[Jy beam$^{-1}$]} & \colhead{[Jy km s$^{-1}$]}
} 
\startdata
$^{12}$CO $J$=2$-$1 & 0.20 & 0.009 & $7.8\pm0.8$  & $17.5\pm1.8$ \\
                    & 0.35 & 0.008 & $7.3\pm0.7$  & $17.5\pm1.8$ \\
$^{12}$CO $J$=3$-$2 & 0.12 & 0.020 & $9.7\pm1.0$  & $37.2\pm3.7$ \\
                    & 0.35 & 0.015 & $10.6\pm1.1$ & $36.0\pm3.6$ 
\enddata
\tablecomments{Properties of the synthesized channel maps at 
native and binned spectral resolutions.  The noise is defined as the RMS 
variation per {\it channel} in line-free regions.  The peak $I_{\nu}$ refers to 
the peak intensity in a channel of fixed width $\Delta v$.  The integrated flux 
densities ($F_{\nu} = \int I_{\nu} dv$) correspond to the sum of the emission 
at all velocities in a $5\arcsec\times5\arcsec$ square box centered on the 
stellar position.  The quoted errors include a $\sim$10\%\ systematic 
uncertainty in the absolute flux scale.}
\end{deluxetable}

\clearpage

\begin{deluxetable}{ll|cccc|ccc}
\tablecolumns{9}
\tablewidth{0pc}
\tablecaption{Model Parameters \label{tab:params1}}
\tablehead{
\colhead{} & \colhead{} & \colhead{fiducial} & \colhead{high-$q$} & \colhead{high-$M_{\ast}$} & \colhead{high-$i$} & \colhead{hot} & \colhead{non-Kep} & \colhead{warp} \\
\colhead{parameter} & \colhead{units} & \colhead{} & \colhead{} & \colhead{} & \colhead{} & \colhead{} & \colhead{} & \colhead{}}
\startdata
$\log{N_{10}}$              & [cm$^{-2}$]   & 19.00   &  19.42     & 19.28   & 19.60   & 19.00   & 18.81 & 18.83   \\
$\gamma$                    & \nodata       & 0.99    &    0.99    & 0.66    & 0.90    & 0.99    & 0.94  & 0.93    \\
$r_c$                       & [AU]          & 28      &   28   & 36      & 24      & 28      & 32    & 33      \\
$T_{10}$ (2$-$1)            & [K]           & 77      &    110     & 75      & 68      & 77      & 100   & 100     \\
\hspace{0.55cm} (3$-$2)     	   & [K]           & 88      &   115      & 94      & 90      & 88      & 99    & 104     \\
$q$ \hspace{0.25cm} (2$-$1) & \nodata        & 0.38    &     0.65    & 0.39    & 0.32    & 0.38    & 0.49  & 0.53    \\
\hspace{0.55cm} (3$-$2)        & \nodata        & 0.44    &    0.65     & 0.51    & 0.49    & 0.44    & 0.49  & 0.53    \\
$\xi$      & [m s$^{-1}$]  & 20      &    20     & 10      & 10      & 20      & 20    & 15      \\
$M_{\ast}$      & [M$_{\odot}$] & 0.8     & 0.8     & 1.5     & 0.8     & 0.8     & 0.8   & 0.8     \\
\hline
$i_{10}$    & [\degr]       & 5.8     &   5.8      & 6.0     & 8.0     & 5.8     & 5.7   & 7.5     \\
$y$            & \nodata       & 0       &     0    & 0       & 0       & 0       & 0     & 0.15    \\
$r_b$                       & [AU]          & \nodata & \nodata & \nodata & \nodata & \nodata & 57    & \nodata \\ 
$x$                         & \nodata       & 0       & 0       & 0       & 0       & 0       & 0.15  & 0       \\
$\delta T$                  & \nodata       & 1       & 1       & 1       & 1       & 3       & 1     & 1 
\enddata
\tablecomments{The parameter values adopted in the modeling analysis in \S 3.2 
and 3.3.  Each column corresponds to a different model type, and each row 
represents a different model parameter (the subscript `10' denotes that 
parameter value at $r = 10$\,AU).  Note that only the ``warp" model has a 
spatially varying disk inclination: in all other cases $i_{10} = i$ at all 
radii, and $y = 0$ by definition (see \S 3.3.3).  The parameter $r_b$ is only 
defined for the ``non-Keplerian" model; in all other cases $x = 0$ (or $f = 1$, 
at all radii; see \S 3.3.2).  The parameter $\delta T$ corresponds to a 
constant scaling of the temperature profile for $r < 4$\,AU in the ``hot" model 
only: all other models have $\delta T = 1$ by definition. }
\end{deluxetable}


\clearpage

\begin{thebibliography}{}
\bibitem[Andrews et al.(2009)]{andrews09} Andrews, S. M., Wilner, D. J., Hughes, A. M., Qi, C., \& Dullemond, C. P. 2009, \apj, 700, 1502
\bibitem[Andrews et al.(2010)]{andrews10} Andrews, S. M., Wilner, D. J., Hughes, A. M., Qi, C., \& Dullemond, C. P. 2010, \apj, 723, 1241
\bibitem[Andrews et al.(2012)]{andrews12} Andrews, S. M., et al. 2012, \apj, 744, 162
\bibitem[Armitage \& Pringle(1997)]{armitage97} Armitage, P. J., \& Pringle, J. E. 1997, \apj, 88, L47
\bibitem[Arnold et al.(2012)]{arnold12} Arnold, T. J., Eisner, J. A., Monnier, J. D., \& Tuthill, P. 2012, \apj, 750, 119
\bibitem[Augereau et al.(1999)]{augereau99} Augereau, J. C., Lagrange, A. M., Mouillet, D., \& M{\'{e}}nard, F. 1999, \aap, 350, L51
\bibitem[Beckwith \& Sargent(1993)]{beckwith93} Beckwith, S. V. W., \& Sargent, A. I. 1993, \apj, 402, 280
\bibitem[Bitsch \& Kley(2011)]{bitsch11} Bitsch, B., \& Kley, W. 2011, \aap, 530, A41
\bibitem[Bouvier et al.(1999)]{bouvier99} Bouvier, J., et al. 1999, \aap, 349, 619
\bibitem[Brinch \& Hogerheijde(2010)]{brinch10} Brinch, C., \& Hogerheijde, M. R. 2010, \aap, 523, A25
\bibitem[Calvet et al.(2002)]{calvet02} Calvet, N., D'Alessio, P., Hartmann, L., Wilner, D. J., Walsh, A., \& Sitko, M. 2002, \apj, 568, 1008
\bibitem[Cassen \& Moosman(1981)]{cassen81} Cassen, P., \& Moosman, A. 1981, Icarus, 48, 353
\bibitem[Chiang \& Murray-Clay(2004)]{chiang04} Chiang, E. I., \& Murray-Clay, R. A. 2004, \apj, 607, 913
\bibitem[Clampin et al.(2003)]{clampin03} Clampin, M., et al. 2003, \aj, 126, 385
\bibitem[Corder et al.(2005)]{corder05} Corder, S., Eisner, J., \& Sargent, A. 2005, \apj, 622, L133
\bibitem[Dawson et al.(2011)]{dawson11} Dawson, R. I., Murray-Clay, R. A., \& Fabrycky, D. C. 2011, \apj, 743, L17
\bibitem[Dutrey et al.(1994)]{dutrey94} Dutrey, A., Guilloteau, S., \& Simon, M. 1994, \aap, 286, 149
\bibitem[Dutrey et al.(1998)]{dutrey98} Dutrey, A., Guilloteau, S., Prato, L., Simon, M., Duvert, G., Schuster, K., \& M{\'{e}}nard, F. 1998, \aap, 338, L63
\bibitem[Dutrey et al.(2008)]{dutrey08} Dutrey, A., et al. 2008, \aap, 490, L15
\bibitem[Evans et al.(2012)]{evans12} Evans, T.~M., Ireland, 
M.~J., Kraus, A.~L., et al.\ 2012, \apj, 744, 120 
\bibitem[Flaherty \& Muzerolle(2010)]{flaherty10} Flaherty, K. M., \& Muzerolle, J. 2010, \apj, 719, 1733
\bibitem[Fukagawa et al.(2004)]{fukagawa04} Fukagawa, M., et al. 2004, \apj, 605, L53
\bibitem[Fukagawa et al.(2006)]{fukagawa06} Fukagawa, M., et al. 2006, \apj, 636, L153
\bibitem[Grady et al.(2001)]{grady01} Grady, C. A., et al. 2001, \aj, 122, 3396
\bibitem[Guilloteau \& Dutrey(1998)]{guilloteau98} Guilloteau, S., \& Dutrey, A. 1998, \aap, 339, 467
\bibitem[Hartmann et al.(1998)]{hartmann98} Hartmann, L., Calvet, N., Gullbring, E., \& D'Alessio, P. 1998, \apj, 495, 385
\bibitem[Heap et al.(2000)]{heap00} Heap, S. R., Lindler, D. J., Lanz, T. M., Cornett, R. H., Hubeny, I., Maran, S. P., \& Woodgate, B. 2000, \apj, 539, 435
\bibitem[Hu{\'e}lamo et al.(2008)]{huelamo08} Hu{\'e}lamo, N., et al. 2008, \aap, 489, L9
\bibitem[Hughes et al.(2007)]{hughes07} Hughes, A. M., Wilner, D. J., Calvet, N., D'Alessio, P., Claussen, M. J., \& Hogerheijde, M. R. 2007, \apj, 664, 536
\bibitem[Hughes et al.(2008)]{hughes08} Hughes, A. M., Wilner, D. J., Qi, C., \& Hogerheijde, M. R. 2008, \apj, 678, 1119
\bibitem[Hughes et al.(2009)]{hughes09} Hughes, A. M., et al. 2009, \apj, 698, 131
\bibitem[Hughes et al.(2011)]{hughes11} Hughes, A. M., Wilner, D. J., Andrews, S. M., Qi, C., \& Hogerheijde, M. R. 2011, \apj, 727, 85
\bibitem[Kalas \& Jewitt(1995)]{kalas95} Kalas, P., \& Jewitt, D. 1995, \aj, 110, 794
\bibitem[Koerner et al.(1993)]{koerner93} Koerner, D. W., Sargent, A. I., \& Beckwith, S. V. W. 1993, Icarus, 106, 2
\bibitem[Koerner \& Sargent(1995)]{koerner95} Koerner, D. W., \& Sargent, A. I. 1995, \aj, 109, 2138
\bibitem[Krist et al.(2000)]{krist00} Krist, J. E., Stapelfeldt, K. R., M{\'{e}}nard, F., Padgett, D. L., \& Burrows, C. J. 2000, \apj, 538, 793
\bibitem[Krist et al.(2005)]{krist05} Krist, J. E., et al. 2005, \aj, 129, 1008
\bibitem[Lin et al.(2006)]{lin06} Lin, S.-Y., Ohashi, N., Lim, J., Ho, P. T. P., Fukagawa, M., \& Tamura, M. 2006, \apj, 645, 1297
\bibitem[Liu(2004)]{liu04} Liu, M. C. 2004, Science, 305, 1442
\bibitem[Lubow \& Ogilvie(2001)]{lubow01} Lubow, S. H., \& Ogilvie, G. I. 2001, \apj, 560, 997
\bibitem[Lynden-Bell \& Pringle(1974)]{lynden-bell74} Lynden-Bell, D., \& Pringle, J. E. 1974, \mnras, 168, 603
\bibitem[Mannings \& Sargent(1997)]{mannings97} Mannings, V., \& Sargent, A. I. 1997, \apj, 490, 792
\bibitem[Markwardt(2009)]{markwardt09} Markwardt, C. B. 2009, in Astronomical Society of the Pacific Conference Series, Vol.~411, Astronomical Data Analysis Software and Systems XVIII, ed.~D.~A.~Bohlender, D.~Durand, \& P.~Dowler, 251
\bibitem[Marzari \& Nelson(2009)]{marzari09} Marzari, F., \& Nelson, A. F. 2009, \apj, 705, 1575
\bibitem[Mouillet et al.(1997)]{mouillet97} Mouillet, D., Larwood, J. D., Papaloizou, J. C. B., \& Lagrange, A. M. 1997, \mnras, 292, 896
\bibitem[Muto et al.(2012)]{muto12} Muto, T., et al. 2012, \apj, 748, L22
\bibitem[Muzerolle et al.(2000)]{muzerolle00} Muzerolle, J., Calvet, N., Brice{\~n}o, C., Hartmann, L.,
\bibitem[Najita et al.(2010)]{najita10} Najita, J. R., et al. 2010, \apj, 712, 274
\bibitem[Nixon \& Pringle(2010)]{nixon10} Nixon, C. J., \& Pringle, J. E. 2010, \mnras, 403, 1887
\bibitem[Papaloizou \& Lin(1995)]{papaloizou95} Papaloizou, J. C. B., \& Lin, D. N. C. 1995, \apj, 438, 841
\bibitem[Pascucci et al.(2011)]{pascucci11} Pascucci, I., et al. 2011, \apj, 736, 13
\bibitem[Pi{\'{e}}tu et al.(2005)]{pietu05} Pi{\'{e}}tu, V., Guilloteau, S., \& Dutrey, A. 2005, \apj, 443, 945
\bibitem[Pontoppidan et al.(2008)]{pontoppidan08} Pontoppidan, K. M., Blake, G. A., van Dishoeck, E. F., Smette, A., Ireland, M. J., \& Brown, J. 2008, \apj, 684, 1323
\bibitem[Qi et al.(2004)]{qi04} Qi, C., et al. 2004, \apj, 616, L11
\bibitem[Qi et al.(2006)]{qi06} Qi, C., et al. 2006, \apj, 636, L157
\bibitem[Quillen(2006)]{quillen06b} Quillen, A. C. 2006, \apj, 640, 1078
\bibitem[Quillen et al.(2006)]{quillen06} Quillen, A. C., Varni{\'e}re, P., Minchev, I., \& Frank, A. 2006, \aj, 129, 2481
\bibitem[Rettig et al.(2004)]{rettig04} Rettig, T. W., Haywood, J., Simon, T., Brittain, S. D., \& Gibb, E. 2004, \apj, 616, L163
\bibitem[Roberge et al.(2005)]{roberge05} Roberge, A., Weinberger, A. J., \& Malumuth, E. M. 2005, \apj, 622, 1171
\bibitem[Rucinski et al.(2008)]{rucinski08} Rucinski, S. M., et al. 2008, \mnras, 391, 1913
\bibitem[Salyk et al.(2007)]{salyk07} Salyk, C., Blake, G. A., Boogert, A. C. A., \& Brown, J. M.\ 2007, \apj, 655, L105
\bibitem[Salyk et al.(2009)]{salyk09} Salyk, C., Blake, G. A., Boogert, A. C. A., \& Brown, J. M. 2009, \apj, 699, 330
\bibitem[Sch{\"o}ier et al.(2005)]{scholier05} Sch{\"o}ier, F.~L., van der Tak, F.~F.~S., van Dishoeck, E.~F., \& Black, J.~H.\ 2005, \aap, 432, 369 
\bibitem[Sellwood(2010)]{sellwood10} Sellwood, J. A. 2010, in Planets, Stars, \& Stellar Systems Vol.~5, ed. G.~Gilmore (arXiv:1006.4855)
\bibitem[Setiawan et al.(2008)]{setiawan08} Setiawan, J., Henning, T., Launhardt, R., M{\"u}ller, A., Weise, P., \& K{\"u}rster, M. 2008, Nature, 451, 38
\bibitem[Shu et al.(2007)]{shu07} Shu, F. H., Galli, D., Lizano, S., Glassgold, A. E., \& Diamond, P. H. 2007, \apj, 665, 535
\bibitem[Shu et al.(2008)]{shu08} Shu, F. H., Lizano, S., Galli, D., Cai, M. J., \& Mohanty, S. 2008, \apj, 682, L121
\bibitem[Simon et al.(2000)]{simon00} Simon, M., Dutrey, A., \& Guilloteau, S. 2000, \apj, 545, 1034
\bibitem[Terebey et al.(1984)]{terebey84} Terebey, S., Shu, F. H., \& Cassen, P. 1984, \apj, 286, 529
\bibitem[Terquem \& Papaloizou(2000)]{terquem00} Terquem, C., \& Papaloizou, J. C. B. 2000, \aap, 360, 1031
\bibitem[Thommes \& Lissauer(2003)]{thommes03} Thommes, E. W., \& Lissauer, J. J. 2003, \apj, 597, 566
\bibitem[van Leeuwen(2007)]{vanleeuwen07} van Leeuwen, F. 2007, \aap, 474, 653
\bibitem[Weidenschilling(1977)]{weidenschilling77} Weidenschilling, S. J. 1977, \mnras, 180, 57
\bibitem[Weinberger et al.(1999)]{weinberger99} Weinberger, A. J., et al. 1999, \apj, 535, L53
\bibitem[Weinberger et al.(2002)]{weinberger02} Weinberger, A. J., et al. 2002, \apj, 566, 409
\bibitem[Wichmann et al.(1998)]{wichmann98} Wichmann, R., Bastian, U., Krautter, J., Jankovics, I., \& Rucinski, S. M. 1998, \mnras, 301, L39 
\end{thebibliography}
\end{document}